\documentclass[10pt,twocolumn,twoside]{IEEEtran}

\usepackage{amsmath,amsfonts, amssymb}
\usepackage{mathabx, mathtools}

\usepackage{array}
\usepackage[caption=false,font=normalsize,labelfont=sf,textfont=sf]{subfig}
\usepackage{textcomp}
\usepackage{stfloats}
\usepackage{url}
\usepackage{verbatim}
\usepackage{cite}
\usepackage[shortlabels]{enumitem}
\usepackage{authblk}
\usepackage{booktabs} 
\usepackage{indentfirst}
\usepackage{times}
\usepackage{nicefrac}
\usepackage{microtype}
\usepackage{mdwmath}

\usepackage{amsthm}
\usepackage{hyperref}
\usepackage[utf8]{inputenc} 
\usepackage[T1]{fontenc}  
\usepackage[ruled]{algorithm2e}
\usepackage{graphicx}
\usepackage{xspace}
\usepackage{cleveref}
\usepackage{tabularx}
\usepackage{ragged2e}
\usepackage[table]{xcolor}
\def\BibTeX{{\rm B\kern-.05em{\sc i\kern-.025em b}\kern-.08em
    T\kern-.1667em\lower.7ex\hbox{E}\kern-.125emX}}

\newtheorem{theorem}{Theorem}
\numberwithin{theorem}{section}
\newtheorem{lemma}{Lemma}
\newtheorem{assumption}{Assumption}
\newtheorem{corollary}{\textit{Corollary}}

\numberwithin{corollary}{section}

\newtheorem{definition}{Definition}
\newtheorem{remark}{Remark}

\newcommand{\IMPC}{{\textsc{InstructMPC}}\xspace}
\newcommand{\removelatexerror}{\let\@latex@error\@gobble}

\begin{document}

\title{Context-Aware Model Predictive Control for Microgrid Energy Management via LLMs}

\author{Ruixiang~Wu$^*$, Jiahao~Ai$^*$, Tinko~Sebastian~Bartels$^*$, Tongxin~Li
\thanks{Ruixiang~Wu, Tinko~Sebastian~Bartels, and Tongxin~Li are with School of Data Science, The Chinese University of Hong Kong, Shenzhen.}
\thanks{Jiahao~Ai is with Department of Statistics and Data Science, the Wharton School, University of Pennsylvania. }
\thanks{$*$ denotes equal contribution.}
\thanks{This work has been submitted to the IEEE for possible publication. Copyright may be transferred without notice, after which this version may no longer be accessible.}
}

\markboth{arXiv Preprint}%
{arXiv Preprint}


\maketitle

\begin{abstract}
    The optimal operation of modern microgrids, particularly those integrating stochastic renewable generation and battery energy storage system~(BESS), relies heavily on load and disturbances forecasting to minimize operational costs. However, in environments with uncertainties in both generation and consumption, traditional numerical forecasting methods often fail to capture generation shifts and event-driven load surges. While contextual information regarding event schedules, system logs, and computational task records is easily obtainable, classic control paradigms lack a formal interface to integrate the unstructured, semantic data into the physical operation loop. This paper addresses this gap by introducing the \IMPC framework, which utilizes a Large Language Model (LLM) paired with a tunable last layer mapping to translate unstructured operational context into predictive disturbance trajectories for the MPC controller. Unlike conventional forecasting methods, the proposed approach treats the last layer mapping as a tunable component, refined online based on the realized control cost. We establish a theoretical foundation for this closed-loop tuning strategy, proving a regret bound of $O(\sqrt{T \log T})$ for linear systems under a tailored task-aware loss function, together with robustness guarantees against uninformative or noisy textual inputs. The control strategy is experimentally validated on OpenCEM, a real-world microgrid with highly fluctuating generation and consumption. Experimental results demonstrate that the LLM-driven MPC significantly reduces cumulative grid electricity costs compared to classical context-agnostic baselines, validating the efficacy of integrating semantic information directly into physical control loops.
\end{abstract}


\section{Introduction}

\IEEEPARstart{E}{nergy} management systems~(EMS) for microgrids, buildings, and industrial sites lie at the center of efforts to integrate stochastic renewable generation, storage, and flexible demand while maintaining reliable grid interaction~\cite{10906491,apablaza2025valuing,10761966, 10529228, 10098893, 10750304}. Operators seek to limit energy cost, reduce unnecessary grid imports, and respect physical limits on batteries and converters. Doing so under uncertainty requires repeated decisions about when to charge or discharge storage, when to import or export power, and how to coordinate flexible loads. A practical bottleneck is the quality of short-horizon predictions of the \emph{net disturbance} that shifts stored energy, meaning the imbalance between uncontrollable generation and load. When that imbalance shifts abruptly because of operational events, predictors that rely only on past meters or rigid numerical features may lag until the effect is already visible in the data stream.

To connect slower planning layers with fast execution, receding-horizon control has become a standard paradigm. Model Predictive Control~(MPC) is widely used in EMS applications because it optimizes storage and grid actions over a finite lookahead using explicit models and forecasted disturbances~\cite{li2021learning,li2025learning, 10098893, 10547193}. Its strength is the systematic tradeoff among competing objectives over time, but its behavior is tightly coupled to the disturbance predictions supplied at each step. Underestimating an upcoming load surge can lead to expensive grid purchases or constraint violations, while systematic forecast bias can strand renewable energy or force curtailment. Closing this gap therefore requires forecasts that reflect not only history but also information that operators and building automation already possess in narrative form.

In real deployments, much of that information appears as \textit{unstructured context}. Maintenance notices, production or event schedules, shift handover notes, work orders, informal operator messages, and text from equipment or IT logs often foreshadow changes in power demand or available generation. These sources are hard to reduce to a small fixed set of numeric tags, yet they are highly correlated with future imbalances. A central question for EMS-oriented MPC is how such semantic information can enter the optimization loop in a disciplined way so that disturbance predictions improve the \emph{downstream} energy management cost, and how the mapping from text to forecasts can be refined online once outcomes are observed. Meeting this goal calls for responsiveness to evolving human input and for learning rules that remain useful when operating conditions drift.

We propose \IMPC, a closed-loop architecture that places a human operator, a Large Language Model~(LLM), and an MPC controller in the same feedback path. Natural-language context is encoded by the LLM, and a \textit{tunable last layer mapping} $g_\theta$ turns the resulting logits into predicted disturbance trajectories that feed standard MPC. After each control interval, the realized disturbance reveals forecast error and drives a task-aware update of $\theta$, so that the semantic interface is judged by control performance rather than by isolated language or pure regression scores alone.

Recent work has extended MPC with richer situational data and with language-based interfaces in several application domains. The energy management literature has particularly emphasized context in structured or numerical form.

\paragraph{Context-aware and data-driven MPC for energy systems} EMS-oriented MPC frequently augments internal models with exogenous numerical signals such as weather, market or tariff indicators, occupancy, and measured states, in order to tighten load and generation forecasts and to avoid overly conservative decisions~\cite{10964552, shi2025disturbanceadaptivedatadrivenpredictivecontrol, ESRAFILIANNAJAFABADI2021107810, DOMA2025115388}. Scenario-based and event-triggered MPC select among predefined policies when labeled events occur~\cite{YANG2023110101, 10922371}. Multi-time-scale schemes combine slower scheduling layers with faster EMS updates driven by new measurements and demand-side programs~\cite{10750304}. These lines of work substantially improve robustness when context is already tabulated or quantized. They leave open the case where the relevant clues appear as free-form text, and they often treat forecasting and control as separate stages so that predictors are not directly adapted using the MPC cost induced by their errors. \IMPC is aimed at closing both gaps for general EMS-style MPC with semantic inputs and closed-loop refinement.

\paragraph{Large language models in smart grids} Large language models are increasingly applied to parse operational narratives, for example in support of situational awareness, event extraction from text, or translation of operator requirements into penalty or constraint templates for learning-based grid tools~\cite{10663471,10675341,11016112}. These contributions improve monitoring, analytics, or offline design of objectives. They do not by themselves provide a formal, repeatable path from unstructured EMS text into the quadratic or economic programs solved at each MPC step, nor do they typically offer guarantees on how online updates behave when text is noisy or irrelevant. Building on our preliminary \IMPC study~\cite{11312156}, this paper develops a coupled LLM-to-MPC pipeline with analysis of online regret and robustness when semantic inputs fail to inform the physical plant. Our contributions are four-fold.
\begin{enumerate}
\item \textbf{Integrated Context-Aware Control Architecture.} We propose \IMPC, a closed-loop framework that formally integrates real-time, unstructured contextual information into the MPC optimization loop. By utilizing an LLM context processor and a tunable last layer mapping, the system enables a classic controller to leverage semantic information, significantly enhancing the adaptability of the system in the presence of complex disturbances.
\item \textbf{Task-Aware Online Learning.} Unlike conventional decoupled forecasting methods, our framework features a task-aware parameter update rule. By fine-tuning the last layer mapping $g_\theta$ based on the realized control cost, we ensure that the LLM produced logits are utilized specifically to minimize downstream control regret, rather than just maximizing linguistic prediction accuracy.
\item \textbf{Convergence and Robustness Guarantees.} We establish a theoretical foundation for the proposed closed-loop tuning. For linear dynamics, we prove a regret bound of $O(\sqrt{T\log T})$ using a tailored loss function. Furthermore, we provide a robustness analysis (Theorem~\ref{thm:robustness}) proving that the system converges to a safe, unbiased baseline even when the contextual inputs are noisy or uninformative, addressing the critical requirement for reliability in control technology.
\item \textbf{Experimental Validation on a Physical Microgrid.} In Section~\ref{sec:experiment}, we experimentally validate the \IMPC framework on OpenCEM~\cite{10.1145/3679240.3734678}, a real-world on-campus PV-and-battery microgrid. The results demonstrate that the LLM-driven controller significantly reduces grid import costs under highly dynamic computing workloads compared to classical context-agnostic baselines.
\end{enumerate}

\section{Preliminaries and Problem Formulation} \label{sec:problem}

This section formalizes the microgrid energy management problem behind \IMPC. We first describe a physical model with photovoltaic~(PV) generation, battery energy storage~(BESS), and controllable grid exchange, then embed it in a linear dynamical system and state the MPC policy. Finally, we introduce contextual information and the online refinement of the mapping from text to disturbance forecasts.

\subsection{Microgrid Energy Management Model} \label{sec:microgrid_model}

We study a microgrid with PV generation, a BESS, and a bidirectional grid interface. Loads vary over time and include both schedulable equipment~(e.g., computing servers, HVAC) and uncontrollable base demand. At each discrete time $t$, energy balance takes the form
\begin{equation}\label{eq:energy_balance}
    s_{t+1} = s_t + u_t + w_t,
\end{equation}
where $s_t \in \mathbb{R}$ is the state-of-charge~(SoC) deviation from a target reference, $u_t \in \mathbb{R}$ is net energy exchanged with the grid~(positive for import, negative for export), and $w_t \in \mathbb{R}$ is the net disturbance, i.e., local PV generation minus load consumption at time $t$. Neither renewable output nor load is perfectly known ahead of time, since PV depends on weather and load depends on occupancy, equipment schedules, and other operational factors.

The EMS objective is to keep battery SoC near a desired level while avoiding excessive grid use. Reference SoC tracking is operationally important because a depleted battery cannot absorb excess PV, whereas a full battery invites costly export or curtailment. We therefore adopt a quadratic cost penalizing SoC deviation and grid exchange,
\begin{equation}\label{eq:energy_cost}
    \sum_{t=0}^{T-1}\big(q\, s_t^2 + r\, u_t^2\big) + p\, s_T^2,
\end{equation}
where $q > 0$ weights SoC deviation, $r > 0$ penalizes grid power~(transaction costs, demand charges, cycling), and $p > 0$ is a terminal penalty on SoC.

\paragraph{Generalization to multi-dimensional systems} The scalar model~\eqref{eq:energy_balance} already reflects the main BESS-centric structure. Richer installations may require a multi-dimensional state, for example multiple storage units, bus voltages, or coupled thermal states. To cover such cases we write the dynamics in linear form. Let $[T]\coloneqq \{0,1,\ldots,T-1\}$. Then
\begin{equation}\label{eq:linear sys}
    x_{t + 1} = A x_t + B u_t + w_t, \quad t\in [T],
\end{equation}
where $x_t \in \mathbb{R}^n$ is the state vector~(e.g., SoC deviations across units), $u_t \in \mathbb{R}^m$ is the control input~(e.g., grid dispatch), and $A \in \mathbb{R}^{n \times n}$, $B \in \mathbb{R}^{n \times m}$ encode topology and storage dynamics. The disturbance $w_t \in \mathbb{R}^n$ aggregates uncontrolled injections and withdrawals~(renewable generation minus load) and is unknown to the controller at time~$t$. Setting $A=I$, $B=I$, and $n=m=1$ recovers~\eqref{eq:energy_balance}.

The controller seeks to minimize the following quadratic cost:
\begin{subequations} \label{eq:costs}
\begin{align}
\label{eq:quadratic_costs}
J^{\star}\coloneqq\min_{(u_t:t\in [T])}&\sum_{t=0}^{T-1}(x_t^{\top}Qx_t+u_t^{\top}Ru_t)+x_T^{\top}Px_T,\\
\label{eq:sys_constraints}
&\text{subject to }\eqref{eq:linear sys},
\end{align}
\end{subequations}
where $Q, R \succ 0$ are positive definite cost matrices~(generalizing the scalar weights $q$ and $r$ in~\eqref{eq:energy_cost}), and $P$ is the solution to the discrete algebraic Riccati equation~(DARE),\footnote{\scriptsize To simplify the presentation of our main results, we fix the terminal cost in~\eqref{eq:quadratic_costs} to be $P$. The arguments extend to more general terminal costs as well, since the overall cost only differs by an $O(1)$ term~\cite{yu2022competitivecontroldelayedimperfect,Li_2022}.}
\begin{equation} \label{eq:dare}
    P = Q+ A^{\top}PA - A^{\top}PB(R+B^{\top}PB)^{-1}B^{\top}PA.
\end{equation}
The disturbance is bounded by a constant $W>0$, i.e., $w_t\in\mathcal{W}\coloneqq \{w:\|w\|\leq W\}$ for all $t\in [T]$, which is natural when PV is capacity-limited and load is bounded by installed ratings.
Furthermore, we invoke a standard assumption on $(A,B)$~(cf.~\cite{dullerud2013course}).
\begin{assumption}\label{asp:mpc}
The system $(A,B)$ is stabilizable.\footnote{\scriptsize In other words, there exists $K\in\mathbb{R}^{m\times n}$ such that $A-BK$ has a spectral radius $\rho$ less than $1$. Thus, the Gelfand’s formula implies that there exist $C>0$, $\rho\in (0,1)$ such that $\|A-BK\|^{t}\leq C\rho^t$ for all $t\geq 0$. }
\end{assumption}

\begin{remark}
While the quadratic cost~\eqref{eq:quadratic_costs} serves as the analytical foundation for our regret analysis, practical microgrid operation often targets direct economic objectives such as minimizing grid electricity procurement costs under time-varying tariffs. In Section~\ref{sec:real_world_exp}, we demonstrate that the \IMPC framework extends naturally to such non-quadratic, price-based objectives.
\end{remark}

\subsection{MPC for Microgrid Energy Management}

Fix a prediction horizon to be an integer $k$ and define  $\mathcal{T} \coloneqq \min\{t + k - 1, T - 1\}$.
We consider the setting in which, at each time $t\in [T]$, $k$-step predictions
$
\hat{w}_{t:\mathcal{T}|t} \coloneqq (\hat{w}_{t|t}, \ldots, \hat{w}_{\mathcal{T}|t})\in\mathcal{W}^k
$ corresponding to future net disturbances~(renewable generation minus load)
are available. Accurate short-term forecasts of these disturbances are critical: underestimating load leads to costly emergency grid imports during peak tariff periods, while overestimating it results in unnecessary battery depletion.

Model predictive control naturally incorporates these predictions into its optimization framework. At each time $t\in [T]$, the controller solves the following optimization problem given the current state $x_t$~(e.g., current battery SoC deviation) and disturbance predictions $\hat{w}_{t:\mathcal{T}|t}$:
\begin{align} \label{eq:mpc_formulation}
    u_{t:\mathcal{T}}^{\textsc{MPC}} &\coloneqq  \arg\min_{u} \Big (\sum_{\tau = t}^{\mathcal{T}} (x_{\tau}^{\top} Q x_{\tau} + u_{\tau}^{\top} R u_{\tau}) + x^{\top}_{\mathcal{T} + 1} P x_{\mathcal{T} + 1}\Big) \notag\\
    & \text{s.t. } x_{\tau + 1} = A x_{\tau} + B u_{\tau} + \hat{w}_{\tau|t}, \forall t \leq \tau < \mathcal{T}.
\end{align}
Only the first element of $u_{t:\mathcal{T}}^{\textsc{MPC}}$ is applied as a control input to the system, and the rest are discarded. This receding-horizon strategy allows the controller to continuously revise its grid exchange plan as new disturbance forecasts become available.

However, in microgrid environments, the net disturbance $w_t$ is notoriously difficult to predict from historical data alone~\cite{binder2019improved, castillo2020predictingfuturestatedisturbed}. Load surges triggered by unscheduled equipment activation, sudden PV ramp-downs due to cloud cover, or demand spikes from large-scale campus events are inherently event-driven and cannot be captured by time-series extrapolation. This motivates incorporating \textit{contextual information}---such as operator schedules, maintenance logs, and event calendars---into the disturbance prediction pipeline.

\subsection{Contextual MPC with Model Fine-Tuning}
Our goal is to improve the performance of the classic MPC defined in~\eqref{eq:mpc_formulation} by incorporating external contextual information available to microgrid operators. At time $t\in [T]$, this contextual information is denoted by $c_{t:\mathcal{T}|t} \coloneqq (c_{t|t}, \ldots, c_{\mathcal{T}|t})$ and encompasses unstructured operational data---such as scheduled maintenance windows, planned computational workloads, weather advisories, and operator directives---that carry predictive value for future net disturbances $\hat{w}_{t:\mathcal{T}|t}\in\mathcal{W}^k$.\footnote{\scriptsize The predicted disturbance trajectory $\hat{w}_{t:\mathcal{T}|t}$ is generated in real-time, operating on the same time scale as the MPC decision-making process.} The central challenge is that this information is \textit{semantic} in nature: it resides in natural language descriptions and system logs rather than in structured numerical time series. To bridge this gap, we integrate an LLM to embed the contextual information into a space of LLM logits $\mathcal{D}$ and map these logits into numerical disturbance predictions useful for downstream MPC. In our framework, the mapping from LLM logits to prediction is modeled by a last layer mapping $g_\theta (\cdot): \mathcal{D} \to \mathcal{W}$, and the parameter $\theta \in \Theta$ is refined as the MPC executes and true disturbances are revealed. Let $\|\cdot\|$ denote the $\ell_2$-norm of a vector. The hypothesis set $\Theta$ has a bounded diameter so that $\|\theta-\theta’\|\leq D$ for all $\theta,\theta’\in \Theta $ and some constant $D>0$. To facilitate a tractable analysis of the online updating process of parameter $\theta$, we impose the following assumption on the structure of $g_\theta$, which serve as approximations for widely used architectures, including transformers~\cite{vaswani2017attention} and input convex neural networks~\cite{amos2017input}.

\begin{assumption}\label{asp:convex}
The last layer mapping $g_\theta$ is differentiable and affine in $\theta$ and its gradient satisfies $\|\nabla_\theta g_\theta\|\leq L$ for some $L>0$.
\end{assumption}
The parameter $\theta_t$ is updated online at each $t \in [T]$ based on realized disturbances, enabling the prediction layer to adapt to evolving operational conditions.
We evaluate performance via the following \emph{regret}, casting the problem as an online optimization over $\theta_{1:T}\coloneqq (\theta_t : t \in [T])$:
\begin{align}
    \label{eq:regret}
    J(\theta_{1:T}) - \min_{\theta \in \Theta} J(\theta),
\end{align}
where $J(\theta_1,\ldots,\theta_{T})$ and $J(\theta^{\star})$ are the quadratic costs defined in~\eqref{eq:quadratic_costs} induced by the MPC actions in~\eqref{eq:mpc_formulation} corresponding to the hyper-parameters $\theta_1,\ldots,\theta_{T}$ and the optimal hyper-parameter $\theta^{\star}$, respectively. In other words, the regret in~\Cref{eq:regret} measures the difference between the cumulative energy management cost incurred by the online sequence $(\theta_t : t \in [T])$, and the cumulative cost achievable by the best fixed parameter $\theta^{\star}$ in hindsight. Next, we introduce the \IMPC framework for contextual MPC.

\section{InstructMPC for Energy Management} 
\label{sec:system}

\begin{figure}[h]
    \centering
    \includegraphics[width =\linewidth]{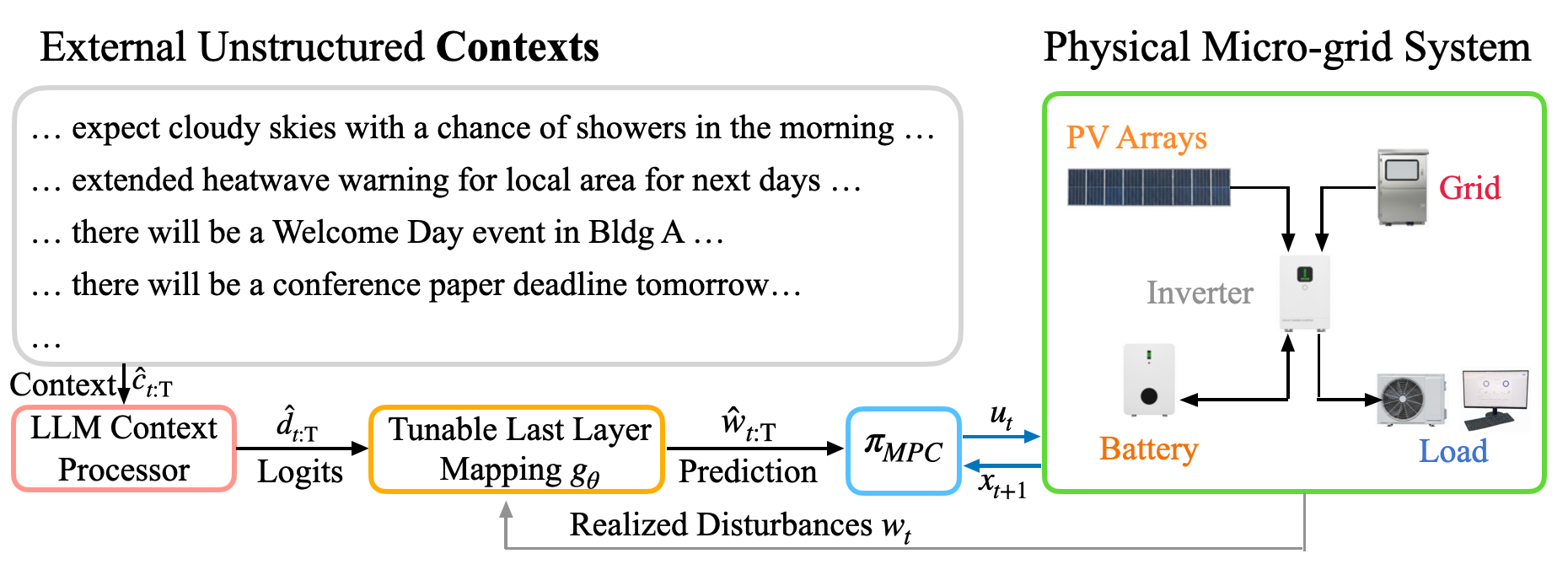}
    \caption{System framework of \IMPC. The blue lines represent interactions between \IMPC and the microgrid, where \IMPC receives the current state $x_t$~(e.g., battery SoC deviation) and outputs the grid exchange decision $u_t$. The black lines represent the \textit{information loop}, within which external contextual information $c_{t: \mathcal{T} | t}$~(e.g., operator schedules, equipment logs) is passed to an LLM together with a last layer mapping to produce predicted net-load disturbances $\hat{w}_{t:\mathcal{T}|t}$. Then, the MPC controller $\pi_{\mathrm{MPC}}$ utilizes $\hat{w}_{t:\mathcal{T}|t}$ and $x_t$ to determine the optimal grid dispatch $u_{t}^{\mathrm{MPC}}$ via~\eqref{eq:mpc_formulation}. After execution, the microgrid reveals the true net disturbance $w_t$~(realized renewable generation minus load). The discrepancy between $w_t$ and $\hat{w}_t$ is then sent back to the last layer mapping $g_\theta$ as a loss signal.}
    \label{fig:system_framework}
        \vspace{-4pt}
\end{figure}

In this section, we introduce \IMPC, a framework that enables MPC-based energy management to leverage unstructured operational context for improved disturbance prediction. In a microgrid setting, this means translating semantic information---such as operator-communicated load schedules, equipment maintenance notices, and event calendars---into quantitative net-load forecasts that the MPC controller can act upon. Fig.~\ref{fig:system_framework} presents the overall architecture of \IMPC. In the following subsections, we explain the core architecture of \IMPC, which combines an LLM with a tunable last layer mapping.
\subsection{Context-Aware Disturbance Prediction via LLMs} \label{sec:cdp}

During the execution of \IMPC, an LLM paired with a tunable last layer mapping predicts future net-load disturbances $\hat{w}_{t:\mathcal{T}|t}$ based on contextual information available at time $t$. The contextual information may include operator-announced schedules~(e.g., ``heavy compute job starting at 6pm''), equipment logs~(e.g., HVAC setpoint changes), or weather-related notices that affect PV output. After each control step, the realized net disturbance $w_t$ is revealed, and the last layer mapping $g_\theta$ refines its parameters $\theta$ by comparing these predictions with the actual observations.

At each time $t$, given contextual information $c_{t:\mathcal{T}|t}$, the LLM outputs a logit vector $d_{\tau|t}$, whose elements directly represent the predicted probabilities of the feature belonging to predefined discrete levels~(e.g., $\{\text{low, medium, high}\}$ load intensity). Finally, the complete logit sequence $\hat{d}_{t : \mathcal{T} | t} \coloneqq (\hat{d}_{t|t}, \ldots, \hat{d}_{\mathcal{T}|t}) \in \mathcal{D}^k$ is fed into the last layer mapping $g_\theta$ to produce the predicted disturbance trajectory $\hat{w}_{t:\mathcal{T}|t}$.

\subsection{\IMPC Framework} 
\begin{algorithm}[h]
\small
\label{alg:main}
\caption{\IMPC for energy management}
\For{$t = 0, \dots, T-1$}{
$\mathcal{T} \leftarrow \min(t + k - 1, T - 1)$\\
\textbf{Get} logits $d_{t:\mathcal{T}|t}$ from LLM with instruction $c_{t:\mathcal{T}|t}$:
\vskip 0.2em
$(\hat{w}_{t|t}, \ldots, \hat{w}_{\mathcal{T}|t})\gets \big(g_{\theta_t}(d_{t|t}), \ldots, (g_{\theta_t}(d_{\mathcal{T}|t})\big)$
    \vskip 0.2em
    Generate an action $u_t$ by solving MPC in~\eqref{eq:mpc_formulation}
    \vskip 0.2em 
    \textbf{Update} 
    \vspace{-4pt}
    \begin{align*}
        x_{t+1} &= A x_t + B u_t + w_t\\
        \theta_{t+1} &= \Pi_\Theta \left( \theta_t-\eta_t\nabla_\theta L_{t-k+1}(\theta_{t-k+1})\right)
    \end{align*}
    \vspace{-4pt}
}
\end{algorithm}

The last layer $g_{\theta}$ is updated at every time $t$ through a loss function $L_{t-k+1}:\Theta\times\mathcal{W}^{2k}\rightarrow\mathbb{R}$, which also depends on realized net disturbances $w_{t:\mathcal{T}}$~(i.e., the actual renewable-minus-load imbalance) and the corresponding predictions $\hat{w}_{t:\mathcal{T}|t}$. An example of such a loss function can be found in Corollary~\ref{coro:final_bound} in Section~\ref{sec:main}. At every time $t \in [T]$, the parameter $\theta$ is updated according to:
\begin{equation} \label{eq:theta_update}
    \theta_{t+1}=\Pi_\Theta \left(\theta_t-\eta_t\nabla_\theta L_{t-k+1}(\theta_{t-k+1})\right),
\end{equation}
where $\eta_t\nabla_\theta L_{t-k+1}$ denotes the gradient of $L_{t-k+1}$ with respect to $\theta$, $\Pi_\Theta$ denotes projection onto the hypothesis set $\Theta$ and $\eta_t$ is a time-varying learning rate. This last layer adaptation technique is widely used to efficiently fine-tune NN models~(see~\cite{galashov2025closed, boix2024towards, taheri2025closingloopinsideneural}). The overall procedure of \IMPC is summarized in Algorithm~\ref{alg:main}. Since $\nabla_\theta L_t(\theta_{t})$ is available only after $(w_t,\ldots,w_{\mathcal{T}})$ are revealed at time $t+k$, a delayed gradient is used in the update rule~\eqref{eq:theta_update}. For convenience, we define $L_{t-k+1}=0$ for $t<k$. In practice, when $t<k$, the NN model does not update due to insufficient observed data.

\section{Theoretical Guarantee}
\label{sec:main}

Having established the microgrid energy management formulation and the \IMPC architecture, we now provide theoretical guarantees on the online learning performance of the last layer mapping. Our main result in this section explores the online optimization of $(\theta_t:t\in [T])$ in the online fine-tuning process. We make the following standard assumption on the loss function $L_t$. Throughout this section, we use $d_\tau$ to denote $d_{\tau|t}$, which is the corresponding component of the sequence $d_{\tau:\mathcal{T}|t}$, omitting the subscript $t$ whenever the time $t$ is clear from the context.

\begin{assumption}\label{asp:lg}
 The gradient $\nabla_\theta L_t(\theta)$ is uniformly bounded, i.e., there exists $G>0$ such that $\|\nabla_\theta L_t(\theta)\|\leq G$ for all $\theta\in\Theta$.
\end{assumption}

A fundamental challenge is that the surrogate loss function used to train $g_\theta$ does not match the true, regret-optimal objective~\eqref{eq:mpc_formulation}, which is unknown at training time. This misalignment creates the primary difficulty, i.e., ensuring that training the upstream decoder module $g_\theta$ actually improves the downstream MPC controller's performance. Our solution is to define and analyze the \textit{loss discrepancy}, a term that quantifies the divergence between the gradient of the decoder module surrogate loss $L$ and that of the true decision loss in~\Cref{eq:mpc_formulation}.


\begin{definition}
\label{def:df}
The loss discrepancy between two  loss functions $L_1(\theta)$ and $L_2(\theta)$, $\emph{\texttt{LD}}(L_1,L_2)$ is defined as 
\begin{equation*}
    \emph{\texttt{LD}}(L_1,L_2) \coloneqq \sup_{\theta \in \Theta}\left\|\nabla_\theta L_1(\theta)-\nabla_\theta L_2(\theta)\right\|.
\end{equation*}
\end{definition}
Next, we will provide the theoretical guarantee of \IMPC from both consistency and robustness aspect.

\paragraph{Consistency of \IMPC Framework}
The theorem below provides a regret bound~(see~\Cref{eq:regret}) for the proposed \IMPC, which interacts with the dynamic system described via~\Cref{eq:linear sys}.

\begin{theorem}\label{thm:main}
    Under Assumption \ref{asp:convex},\ref{asp:lg}, if the learning rate $\eta_t$ is non-increasing, then
\begin{align}
\nonumber
    J(\theta_{1:T}) -& J(\theta^{\star})
    \leq \frac{D^2}{2\eta_{T-1}}+\left(k-\frac{1}{2}\right)G^2\sum_{t=0}^{T-1}\eta_t\\
    \label{eq:loss_discrepancy_bound}
    &+D\sum_{t=0}^{T-1}\emph{\texttt{LD}}(L_t,\psi_t^\top H\psi_t)+(k-1)GD.
\end{align}
Furthermore, if we choose $\eta_t=\frac{D}{G\sqrt{2(2k-1)(t+1)}}$,
\begin{align*}
    J(\theta_{1:T})-&J(\theta^{\star})\leq 2GD\sqrt{(k-\frac{1}{2})T}\\
    &+D\sum_{t=0}^{T-1}\emph{\texttt{LD}}\left(L_t,\psi_t^\top H\psi_t\right)+(k-1)GD,
\end{align*}
where $H$ is defined in Lemma~\ref{lem:regret}, and we define
\begin{equation*}
    \psi_t(\theta) \coloneqq \sum_{\tau=t}^{T-1}\left(F^\top\right)^{\tau-t}Pw_\tau-\sum_{\tau=t}^{\mathcal{T}}\left(F^\top\right)^{\tau-t}Pg_{\theta_t}(d_{\tau}).
\end{equation*}
\end{theorem}

\begin{proof}
    
By Theorem 3.2 in~\cite{yu2021powerpredictionsonlinecontrol}, given predictions $\hat{w}_{t:\mathcal{T}|t}$, the MPC 
 solution $u_t^{\textsc{MPC}}$ to the problem defined in~\eqref{eq:mpc_formulation} is
    \begin{equation*}
        u_t^{\textsc{MPC}} = -(R+B^{\top}PB)^{-1} B^{\top}\Big(PAx_t + \sum_{\tau = t}^{\mathcal{T}} (F^{\top})^{\tau - t} P \hat{w}_{\tau|t}\Big ),
    \end{equation*}
where $F \coloneqq A - B(R + B^{\top}PB)^{-1}B^{\top}PA \coloneqq A - BK$.
Thus, applying Lemma \ref{lem:regret} twice, we obtain
    \begin{equation*}
        J(\theta_{1:T})-J(\theta^{\star})=\sum_{t=0}^{T-1}\psi_t(\theta_t)^\top H\psi_t(\theta_t)-\psi_t(\theta^{\star})^\top H\psi_t(\theta^{\star}),
    \end{equation*}
where $\psi_t(\theta)$ is defined in Theorem~\ref{thm:main}. For notational simplicity, we define $\psi_t\coloneqq \psi_t(\theta_t)$ and $\phi_t\coloneqq \psi_t(\theta^{\star})$.
By our model assumption, $Q,R\succ 0$ implies $H\succeq 0$, using the convexity of $x^\top Hx$,
\begin{align} 
    &\frac{1}{2}\sum_{t=0}^{T-1}\psi_t^\top H\psi_t-\phi_t^\top H\phi_t \leq\sum_{t=0}^{T-1}\psi_t^\top H(\psi_t-\phi_t)  =\nonumber\\
    \nonumber
    &\sum_{t=0}^{T-1}\psi_t^\top H \left(\sum_{\tau=t}^{\mathcal{T}}\left(F^\top\right)^{\tau-t}P\left(g_{\theta^{\star}}(d_{\tau})-g_{\theta_t}(d_{\tau})\right)\right).
\end{align} 

Applying Assumption~\ref{asp:convex}, since $g_{\theta}$ is affine in $\theta$, continuing from above,
\begin{align}
\nonumber
&J(\theta_{1:T})-J(\theta^{\star})\leq\\
&-2\sum_{t=0}^{T-1}\psi_t^\top H \left(\sum_{\tau=t}^{\mathcal{T}}\left(F^\top\right)^{\tau-t}P\nabla_\theta g_{\theta_t}(d_\tau)^\top\right)(\theta_t-\theta^{\star})\notag \\
&=\sum_{t=0}^{T-1}\nabla_{\theta_t}(\psi_t^\top H \psi_t)^\top(\theta_t-\theta^{\star}),
\label{eq:linear regret}
\end{align} 
where we denote $\nabla_{\theta_t}(\psi_t^\top H \psi_t)\coloneqq\nabla_{\theta_t}(\psi_t^\top H \psi_t)|_{\theta=\theta_t}$, and have used Assumption \ref{asp:convex}. 
Denote the gradient of the loss function as $\nabla_\theta L_{t-k+1}(\theta_{t-k+1})$ as $l_{t-k+1}$. It follows that the RHS of~\eqref{eq:linear regret} can be rewritten as 
\begin{align}
\nonumber
&\sum_{t=0}^{T-1}\nabla_{\theta_t}(\psi_t^\top H \psi_t)^\top(\theta_t-\theta^{\star}) \notag\\
        =&\sum_{t=0}^{T-1}l_{t-k+1}^\top(\theta_t-\theta^{\star})+\sum_{t=0}^{T-1}(l_t-l_{t-k+1})^\top(\theta_t-\theta^{\star}) \notag\\
        &\quad + \sum_{t=0}^{T-1}(\nabla_{\theta_t}(\psi_t^\top H \psi_t)^\top-l_t^\top)(\theta_t-\theta^{\star}).\label{eq:decomposed_regret}
\end{align}
The bound in~\eqref{eq:loss_discrepancy_bound} is obtained by bounding the terms in~\Cref{eq:decomposed_regret} separately, and the details are relegated to Appendix~\ref{app:proof_main}.
\end{proof}

Furthermore, for a wide range of loss functions $(L_t:t\in [T])$, as exemplified in Corollary \ref{coro:final_bound}, they exhibit a bounded discrepancy from the MPC cost with the discrepancy decaying exponentially as $k$ increases, i.e., $\emph{\texttt{LD}}(L_t,\psi_t^\top H\psi_t)\leq C\rho^k$. 

\begin{corollary}\label{coro:final_bound} Under Assumption \ref{asp:mpc}, for the case when $L_t(\theta)=\hat{\psi}_t(\theta)^\top H\hat{\psi}_t(\theta)$ with
\[
    \hat{\psi}_t(\theta) \coloneqq \sum_{\tau=t}^{\mathcal{T}}\left(F^\top\right)^{\tau-t}Pw_\tau-\sum_{\tau=t}^{\mathcal{T}}\left(F^\top\right)^{\tau-t}Pg_{\theta_t}(d_\tau),
\]
there exist constants $C$ and $0<\rho<1$ such that 
    \[
    \emph{\texttt{LD}}(L_t,\psi_t^\top H\psi_t)\leq C\rho^k, \ \text{ for all } t\in [T],
    \]
    where
   $\psi_t(\theta)$ is defined in Theorem~\ref{thm:main}.
\end{corollary}

For the particular choice of $L_t$ in Corollary~\ref{coro:final_bound}, the term 
$
D\sum_{t=0}^{T-1}{\texttt{LD}}(L_t,\psi_t^\top H \psi_t)
$
in the bound~\eqref{eq:loss_discrepancy_bound} of Theorem~\ref{thm:main} simplifies to $CDT\rho^k$. This provides an insightful guideline for selecting $k$. Specifically, by setting 
$
k = \log_{1/\rho} T,
$
we can achieve a regret bound of $O(\sqrt{T\log T})$. 

Given the sub-linear regret bound guarantees performance under general conditions, we also need to ensure the \textit{robustness} to low-quality inputs in practical power system operations. In real-world microgrid operations, the upstream LLM or encoder might fail to extract meaningful information from the provided context---for instance, when operator logs are uninformative or system events are unrelated to the electrical load---and yield low-quality logits. Next, we provide the robustness guarantee of \IMPC framework. 

\paragraph{Robustness of \IMPC Framework} We will analyze the robustness result in a case, where environmental disturbance $w_t$ and the corresponding context $c_t$ are sampled independently and identically distributed from a fixed joint distribution. Consequently, the resulting LLM logits $d_t$ are also treated as stochastic variables. First, to ensure that the convergence result is theoretically tractable, we impose that the optimal solution lies within the interior of the hypothesis set. Second, since the context and logits are random variables, we further assume that the LLM logits are pair-wise independent, which incurs no loss of generality, as we can always restrict our attention to a basis of the feature space. Third, to simplify the presentation, we consider an over-actuated system $(A, B)$ with some full rank matrix $B$. It guarantees that the controller possess full authority to correct disturbances in any direction. Finally, existing literature~\cite{pmlr-v28-shamir13} suggests that, when using gradient descent, the distance between the current parameter and the optimal one shrinks proportionally to the inverse of time (e.g., $O(1/t)$ or $O(1 / \sqrt{t})$). Therefore, we assume a sufficiently large horizon to ensure the online updates have adequate time to converge.

\begin{assumption} \label{asp:SGD_robustness}
We need the following assumptions:
\begin{enumerate}[ref=\theassumption (\arabic*)]
\item \label{asp:SGD_robustness_feasible}
    The unconstrained minimizer of the loss in Corollary~\ref{coro:final_bound} lies in the interior of the hypothesis set $\Theta$;
\item \label{asp:SGD_robustness_independent_feature}
    The entries of the LLM logits $d$ are pair-wise independent;
\item \label{asp:SGD_robustness_overactuated}
    The system matrix $B \in \mathbb{R}^{n \times m}$ has full row rank;
\item \label{asp:SGD_robustness_infinite_horizon}
    The control task horizon $T$~(see~\eqref{eq:linear sys}) is large enough.
\end{enumerate}
\end{assumption}

To theoretically guarantee robustness, we consider a practical last layer mapping $g_{\theta_t} (d) = C_t d + b_t$, where $\theta_t = [C_t, b_t]^\top$. It serves as a first-order approximation of the NN's behavior and allows us to derive explicit convergence limits.
\begin{theorem}\label{thm:robustness} Under Assumption~\ref{asp:lg} and~\ref{asp:SGD_robustness}, consider the affine mapping defined above. If the LLM logits $d$ are statistically uncorrelated with the disturbances $w$, then for the loss function $L_t(\theta)$ defined in Corollary~\ref{coro:final_bound}, the sequence of parameters $\{\theta_t\}_{t \in [T]}$ produced by Algorithm~\ref{alg:main} satisfies
$
    \lim_{t \to \infty} C_t = 0$ and $\lim_{t \to \infty} b_t = \mathbb{E}[w],
$
where $\mathbb{E}[w]$ is the expected value of the disturbance.
\end{theorem}

Theorem~\ref{thm:robustness} provides a concrete, practical interpretation of robustness within the \IMPC framework. In real-world operations, if the LLM processes uninformative context or generates hallucinated, uncorrelated logits (i.e., the outputs $d$ have no statistical correlation with the true physical disturbances $w$), the task-aware online updating mechanism actively penalizes reliance on the LLM. Mathematically, $\lim_{t \to \infty} C_t = 0$ indicates that the weight matrix assigned to the LLM logits decays to zero, effectively severing the unreliable semantic information from the control loop. Simultaneously, $\lim_{t \to \infty} b_t = \mathbb{E}[w]$ ensures that the predictor falls back to a safe, baseline statistical estimator. In other words, the disturbance prediction $g_{\theta_t}(d)$ eventually depends entirely on the historical statistics (the mean) of the physical disturbances $w$, completely bypassing the faulty LLM outputs. This mechanism guarantees that the \IMPC framework remains stable and reliable, avoiding catastrophic control failures even when context processors fail.

\section{Numerical Results} \label{sec:experiment}

\begin{figure}[ht]
    \centering
    \vspace{2mm}\includegraphics[width=0.95\linewidth]{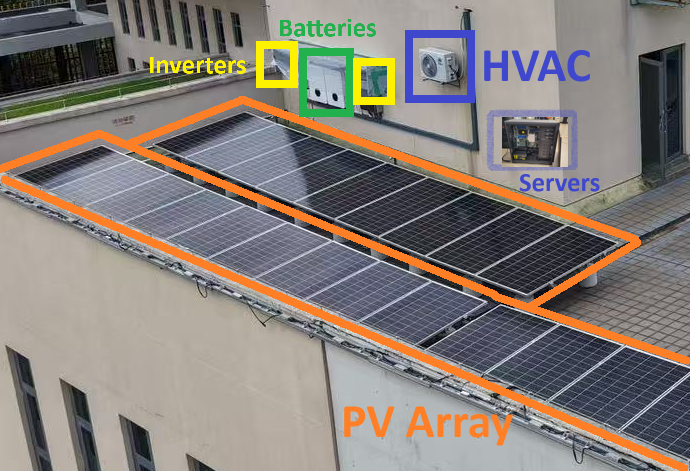}
    \caption{\textbf{OpenCEM on-campus installation.}}
    \label{fig:opencem_installation}
\end{figure}

We validate the efficacy of \IMPC framework through two applications that bridge the gap between control theory and practical power systems operation. We first conduct a theoretic validation in Subsection~\ref{sec:soc_tracking}, where we focus on State-of-Charge~(SoC) tracking under idealized linear dynamics. This scenario is specifically designed to align with the problem formulation in Section~\ref{sec:problem}, allowing us to empirically verity the sublinear regret bounds derived in Corollary~\ref{coro:final_bound}. In Subsection~\ref{sec:real_world_exp}, we move beyond artificial reference tracking task to practical operational cost minimizating under non-linear system dynamics governed by OpenCEM simulator~\cite{bartels2026bridgingnaturallanguagemicrogrid}. By varying the density of available semantic context, we demonstrate the impact of context availability on \IMPC framework. Both experiments are grounded in the physical parameters of the OpenCEM~\cite{10.1145/3679240.3734678} testbed. As shown in Fig.~\ref{fig:opencem_installation}, the installation includes a photovoltaic~(PV) array of $26$ solar panels~($480$ W) for power generation, two $10$ kWh lithium-ion storage, and powers a combined load of high-performance-computing~(HPC) servers and HVAC systems. For prediction, we utilize contextual information extracted from HPC servers' system logs and human-input scheduling information, see Table~\ref{tab:event-records} for examples.

\definecolor{tblstripe}{gray}{0.97}
\definecolor{tblhead}{HTML}{E4EDF5}
\definecolor{srcteam}{HTML}{E5F0E8}
\definecolor{srclog}{HTML}{EFEFEF}
\begin{table*}[t!]
\centering
\caption{Representative unstructured context strings used as inputs $c_{t:\mathcal{T}|t}$ to the LLM in \IMPC (OpenCEM testbed). Logits are mapped through $g_\theta$ to net-disturbance forecasts $\hat{w}_{t:\mathcal{T}|t}$ for MPC\,\protect\eqref{eq:mpc_formulation}. Categories indicate how each snippet informs anticipated computing load and hence the renewable-minus-load imbalance~$w_t$.}
\label{tab:event-records}
\footnotesize
\setlength{\tabcolsep}{5.5pt}
\renewcommand{\arraystretch}{1.2}
\begin{tabularx}{\textwidth}{@{} >{\RaggedRight\arraybackslash}p{2.5cm} >{\RaggedRight\arraybackslash}p{3.2cm} >{\RaggedRight\arraybackslash}X >{\centering\arraybackslash}p{0.8cm} @{}}
\toprule[1pt]
\rowcolor{tblhead}
\textbf{Category} & \textbf{Active interval} & \textbf{Context excerpt (natural language)} & \textbf{Src.} \\
\midrule[0.6pt]
\rowcolor{tblstripe}
Asset metadata & 09-13 23:00--23:20 & Hardware configuration: \texttt{CPU}: Intel Xeon Gold 6526Y; \texttt{GPU}: NVIDIA RTX 2000; additional specs truncated. & \cellcolor{srcteam}{Team} \\
\rowcolor{white}
Announced workload & 07-28 18:00 -- 07-29 16:00 & Tomorrow I will run a CPU-intensive, multi-core numeric robustness test for a day. & \cellcolor{srcteam}{Team} \\
\rowcolor{tblstripe}
Announced workload & 07-29 17:00 -- 07-31 15:00 & In the evening, I will run a CPU-intensive, multi-core numeric robustness test for two days. & \cellcolor{srcteam}{Team} \\
\rowcolor{white}
Transient event & 07-31 04:28--08:00 & Unexpected reboot. & \cellcolor{srclog}{Log} \\
\rowcolor{tblstripe}
Build / CLI activity & 07-28 18:11--19:11 & Shell: \texttt{cd dev/boost/libs} (\ldots). & \cellcolor{srclog}{Log} \\
\rowcolor{white}
Build / CLI activity & 07-28 18:26--19:26 & Shell: \texttt{b2 -j8 cxxflags='-O2'} in \texttt{\ldots/geometry/test}. & \cellcolor{srclog}{Log} \\
\bottomrule[1pt]
\vspace{0.1pt}
\end{tabularx}
{\scriptsize \textit{Note.} ``Team'' denotes user-entered schedules; ``Log'' denotes automated system capture. Internal ingest timestamps omitted for space. Data collected in 2025.}
\vspace{-5pt}
\end{table*}

\subsection{Experiment 1: SoC Tracking under Idealized Dynamics} \label{sec:soc_tracking}

\textbf{Problem description.} We first examine a battery management task where the system is required to track an predefined SoC trajectory. This setup is designed to perfectly align with our problem formulation in~\eqref{eq:costs} with:
\begin{equation*}
    x_{t + 1} = x_t + u_t + w_t, \text{ and }Q = 10^{-4}, P = 10^{-2},
\end{equation*}
where $x_t$ is the deviation from the target SoC, $u_t$ is the net energy bought from the grid (negative if energy is fed back to the grid in timestep $t$), and $w_t$ is the difference of energy generated from solar power and energy consumed by the workstations. The planning horizon is three days, divided into $T=2160$ two-minute steps.

\begin{figure}[h]
    \centering
    \includegraphics[width=0.99\linewidth]{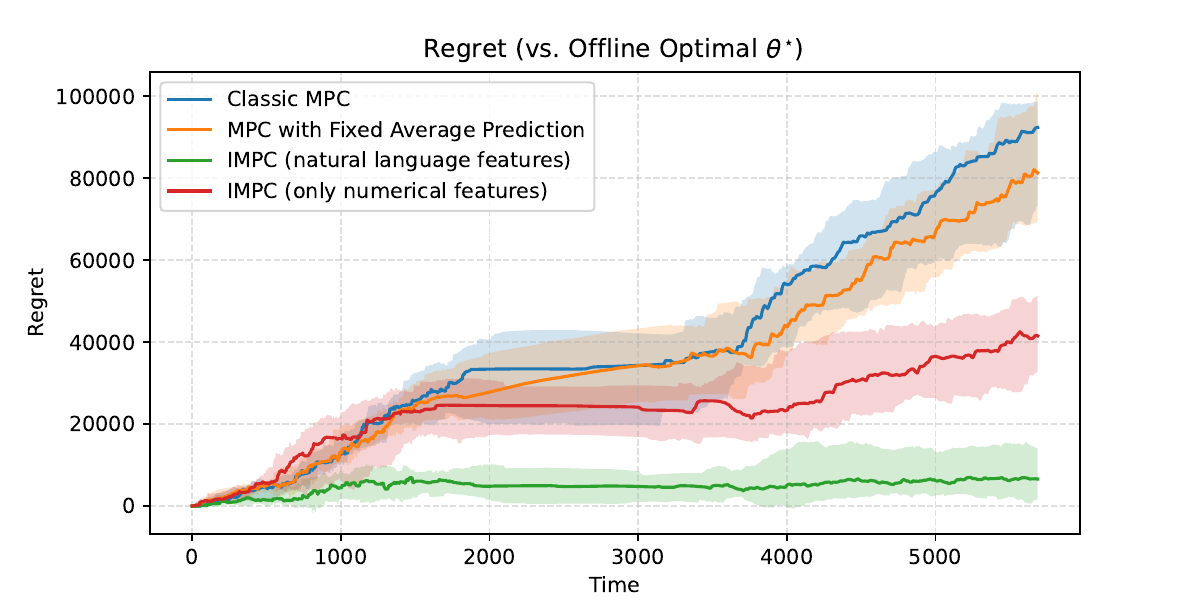}
    \caption{\textbf{Regret Comparison for SoC Tracking (Application 1).} The regret of \IMPC with natural language features~(green) grows sublinearly, empirically confirming the theoretical guarantees provided in Section~\ref{sec:main}. Baselines failing to integrate semantic context exhibit linear regret growth. For each method, we ran $20$ episodes, and plot the average regret trajectory. The shadow region denotes the variance of trajectories.}
    \label{fig:regret_comparison}
    \vspace{-7pt}
\end{figure}

\textbf{Experimental results.} We evaluate the performance of \IMPC via the regret defined in~\eqref{eq:regret}, which compares the online cumulative cost against that of the optimal parameter $\theta^\star$ in hindsight. As illustrated in Fig.~\ref{fig:regret_comparison}, the \IMPC framework utilizing natural language features~(green line) exhibits a clear \textit{sublinear regret} growth. This empirically validates the $O(\sqrt{T \log T})$ bound proven in Corollary~\ref{coro:final_bound}. In contrast, the "Classic MPC baseline"~(blue line) and "MPC with Fixed Average Prediction"~((orange line) exhibit significantly higher, rough linear regret, as they lack the mechanism to adapt to the contextual information. Furthermore, while the \IMPC with only numerical features~(red) reduces the regret compared to baselines, it remains notably higher than \IMPC with natural language features. This highlights that the LLM logits capture critical operational nuances that are omitted in structured numerical metadata.

\subsection{Experiment 2: Operational Cost Minimization in OpenCEM} \label{sec:real_world_exp}

\textbf{Problem description.} This section generalizes the \IMPC framework to a more realistic operational environment. Moving beyond the idealized linear-quadratic assumptions of the fist experiment, this scenario demonstrates the effectiveness of integrating semantic context directly into a complex decision-making loop. The system dynamics is governed by a physics-grounded dygital twin OpenCEM simulator that incorporates practical operational factors. The primary objective is shifted from artificial trajectory tracking to the minimization of actual grid electricity purchasing costs:
\begin{equation*}
    \text{Cost} = \sum_{t=0}^{T-1} p_t \cdot \max(0, u_t),
\end{equation*}
where $p_t$ represents the time-varying electricity price and $u_t$ denotes the net energy imported from the grid at time $t$. By utilizing contextual information exemplified in Table~\ref{tab:event-records}, the \IMPC framework proactively manage the battery state to avoid importing power during expensive peak hours.

\begin{figure*}
    \centering
    \includegraphics[width=0.9\linewidth]{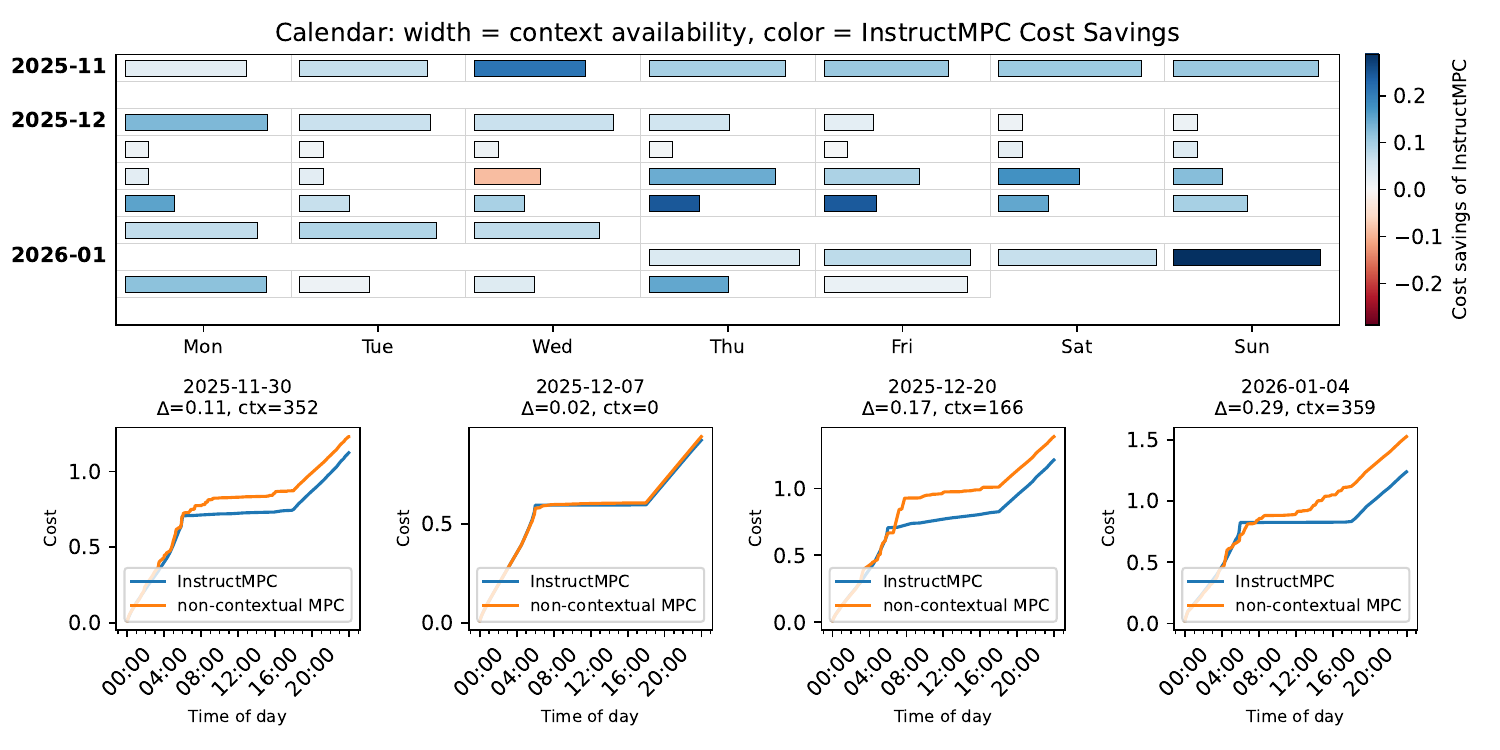}
    \caption{\textbf{Real Physical Microgrid Operational Cost Minimization (Application 2).} The calender heatmap~(Top) illustrates the relationship between daily context availability~(encoded by block width) and the resulting cost savings of \IMPC relative to classic MPC baseline~(encoded by color intensity). The daily cumulative cost profile~(Bottom) compares \IMPC~(blue) with classic MPC~(non-contextual) baseline~(orange). }  
    \label{fig:operetional_cost_minimization}
    \vspace{-4pt}
\end{figure*}

\textbf{Experimental results.} The performance of \IMPC framework was evaluated against a classic MPC~(non-contextual) baseline within the OpenCEM simulator~\cite{bartels2026bridgingnaturallanguagemicrogrid}. The results, visualized in Fig.~\ref{fig:operetional_cost_minimization}, demonstrates that our framework successfully generalized to realistic environments characterized by non-linear dynamics and non-quadratic costs. The insight of this experiment is the \textit{Context-Savings Correlation}. The calender view~(Top of Fig.~\ref{fig:operetional_cost_minimization}) reveals a positive correlation between context availability~(\texttt{ctx}) and cost reduction rate~($\Delta$). On days with dense semantic information, such as~\textsc{2026-01-04}~(\texttt{ctx} $= 359$), our framework achieves a high cost saving of $\Delta = 0.29$. This daily profile indicate that the controller utilizes these high-density instructions to accurately anticipate workstation surges and avoid grid energy procurement during expansive peak-hours.

\section{Conclusion}

We have presented the \IMPC framework that integrates real-time human instructions into Model Predictive Control (MPC) via a human-LLM-in-the-loop approach. This work demonstrates how leveraging contextual information can significantly improve MPC’s predictive accuracy and performance in complex, real-world power system applications. By dynamically generating context-aware disturbance predictions and refining them using a last layer adaptation technique, \IMPC offers improved adaptability and performance across a wide range of applications. For linear dynamics, we established a performance guarantee, proving that \IMPC achieves a regret bound of $O(\sqrt{T\log T})$. Furthermore, we theoretically provides a critical safety guarantee that in the presence of uninformative context or "hallucinated" LLM outputs, the framework's online updates automatically penalize the unreliable semantic information. Future research will explore robust mechanisms to verify LLM outputs against physical grid constraints before they influence the control loop. Additionally, we aim to explore how multiple local controllers, each equipped with its own context-aware predictor, can coordinate to maintain system-wide stability while interpreting localized, heterogeneous human instructions.

\bibliography{biblography}
\bibliographystyle{IEEEtran}


\makeatletter
\renewcommand\appendix{
  \par
  \setcounter{section}{0}
  \setcounter{subsection}{0}
  \gdef\thesection{\@Alph\c@section}
}
\makeatother

\appendix
 
\vspace{-10pt}

\section{Proof of Theorem~\ref{thm:main}}
\label{app:proof_main}


The following lemma characterizes the quadratic cost gap.

\begin{lemma}[Lemma 13 in~\cite{yu2022competitivecontroldelayedimperfect}]\label{lem:regret}
For any $\psi_t\in \mathbb{R}^n$, if at each time $t\in [T]$, a controller $\pi$ implements a control input
$
        u_t^{\pi}=-(R+B^\top PB)^{-1}B^\top\big(PAx_t+\sum_{\tau=t}^{T-1}\left(F^\top\right)^{\tau-t}Pw_\tau-\psi_t\big),
$
    then the difference between the optimal cost $J^{\star}$ and the algorithm cost $J(\pi)$ is given by $
       J(\pi)-J^{\star} = \sum_{t=0}^{T-1}\psi_t^\top H\psi_t,$
where $H \coloneqq B(R+B^\top PB)^{-1}B^\top$ and $F \coloneqq A-HPA$.
\end{lemma}

Now, we bound the terms in~\Cref{eq:decomposed_regret} separately:
\begin{align}
\nonumber
&\sum_{t=0}^{T-1}\nabla_{\theta_t}(\psi_t^\top H \psi_t)^\top(\theta_t-\theta^{\star}) \notag\\
        =&\underbracket{\sum_{t=0}^{T-1}l_{t-k+1}^\top(\theta_t-\theta^{\star})}_{\text{(a)}}+\underbracket{\sum_{t=0}^{T-1}(l_t-l_{t-k+1})^\top(\theta_t-\theta^{\star})}_{\text{(b)}} \notag\\
        &\quad + \sum_{t=0}^{T-1}(\nabla_{\theta_t}(\psi_t^\top H \psi_t)^\top-l_t^\top)(\theta_t-\theta^{\star}).
        \nonumber
\end{align}
First, based on the gradient update rule in~\eqref{eq:theta_update} of \IMPC,
 \begin{align}
&\|\theta_{t+1}-\theta^{\star}\|^2
= \|\Pi_\Theta\left(\theta_t-\eta_t l_{t-k+1}\right)-\theta^{\star}\|^2 \notag\\
\leq& \|\theta_t-\eta_t l_{t-k+1}-\theta^{\star}\|^2 \notag\\
= & \|\theta_t-\theta^{\star}\|^2-2\eta_t l_{t-k+1}^\top(\theta_t-\theta^{\star})+\eta_t^2\|l_{t-k+1}\|^2.
\label{eq:bound_theta_1}
\end{align}
Then, rearranging the terms in~\Cref{eq:bound_theta_1} and noting that the gradients are bounded by Assumption~\ref{asp:lg},
    \begin{align*}
        2l_{t-k+1}^\top(\theta_t-\theta^{\star})\leq \frac{\|\theta_t-\theta^{\star}\|^2-\|\theta_{t+1}-\theta^{\star}\|^2}{\eta_t}+\eta_tG^2.
    \end{align*}
Summing above from $t=0$ to $T-1$, for (a),
    \begin{align}
        2&\sum_{t=0}^{T-1}l_{t-k+1}^\top(\theta_t-\theta^{\star}) \notag\\
        \leq &\sum_{t=0}^{T-1}\frac{\|\theta_t-\theta^{\star}\|^2-\|\theta_{t+1}-\theta^{\star}\|^2}{\eta_t}+\eta_tG^2 \notag\\
        \leq &\sum_{t=0}^{T-1}\|\theta_t-\theta^{\star}\|^2\left(\frac{1}{\eta_t}-\frac{1}{\eta_{t-1}}\right)+G^2\sum_{t=0}^{T-1}\eta_t \notag\\
        \leq &\frac{D^2}{\eta_{T-1}}+G^2\sum_{t=0}^{T-1}\eta_t .\label{eq: bound1}
    \end{align}

Furthermore, the second term in~\Cref{eq:decomposed_regret} can be bounded from above as
    \begin{align}
      \text{(b)}=  &\sum_{t=0}^{T-1}(l_t-l_{t-k+1})^\top(\theta_t-\theta^{\star}) \notag\\
      =&\sum_{t=0}^{T-k}l_t^\top(\theta_t-\theta_{t+k-1})+\sum_{t=T-k+1}^{T-1}l_t^\top(\theta_t-\theta^{\star}) \notag\\
        \leq &\sum_{t=0}^{T-k}(G\|\theta_t-\theta_{t+k-1}\|) +(k-1)GD \notag\\
        =&(k-1)GD+G\sum_{t=0}^{T-k}\left\|\sum_{\tau=t}^{t+k-2}\eta_{\tau}l_{\tau-k+1}\right\| \notag\\
        \leq& (k-1)GD+G^2\sum_{t=0}^{T-k}\sum_{\tau=t}^{t+k-2}\eta_{\tau} \notag\\
        \leq &(k-1)GD+(k-1)G^2\sum_{t=0}^{T-1}\eta_t. \label{eq:bound2}
    \end{align}
Recalling \eqref{eq:linear regret}, 
    \begin{align}
        &J(\theta_{1:T})-J(\theta^{\star}) \notag\\
        \leq &\frac{D^2}{2\eta_{T-1}}+\frac{G^2}{2}\sum_{t=0}^{T-1}\eta_t +(k-1)GD+(k-1)G^2\sum_{t=0}^{T-1}\eta_t \notag\\
        +&\sum_{t=0}^{T-1}\|\nabla_{\theta_t}(\psi_t^\top H \psi_t)-l_t\|\|\theta_t-\theta^{\star}\| \label{eq:-1}\\ 
        \leq &\frac{D^2}{2\eta_{T-1}}+(k-\frac{1}{2})G^2\sum_{t=0}^{T-1}\eta_t+(k-1)GD \notag\\
        +&D\sum_{t=0}^{T-1}\emph{\texttt{LD}}(L_t,\psi_t^\top H\psi_t),\label{eq:fin}
    \end{align}
where we have used~\eqref{eq: bound1} and \eqref{eq:bound2} to derive~ \eqref{eq:-1}; the last inequality~\eqref{eq:fin} holds by the loss discrepancy in~\Cref{def:df}. Finally, applying $\eta_t={D}/\left({G\sqrt{2(2k-1)(t+1)}}\right)$ to \eqref{eq:fin},
    \begin{align}
        &J(\theta_{1:T})-J(\theta^{\star}) \notag\\
        = &GD\sqrt{(k-\frac{1}{2})T} +\sqrt{k-\frac{1}{2}}\frac{GD}{2}\sum_{t=0}^{T-1}\frac{1}{\sqrt{t+1}}\notag\\
        +&D\sum_{t=0}^{T-1}\emph{\texttt{LD}}(L_t,\psi_t^\top H\psi_t)+(k-1)GD \notag\\
        \leq &2GD\sqrt{(k-\frac{1}{2})T}+D\sum_{t=0}^{T-1}\emph{\texttt{LD}}(L_t,\psi_t^\top H\psi_t)+(k-1)GD. \nonumber
    \end{align}
The inequality above in the second statement of Theorem~\ref{thm:main} is obtained by using $\sum_{t=1}^T\frac{1}{\sqrt{t}}\leq 2\sqrt{T}$.

\section{Proof of Corollary~\ref{coro:final_bound}}

Suppose $L_t(\theta)=\hat{\psi}_t(\theta)^\top H\hat{\psi}_t(\theta)$ where
\[
    \hat{\psi}_t(\theta) \coloneqq \sum_{\tau=t}^{\mathcal{T}}\left(F^\top\right)^{\tau-t}Pw_\tau-\sum_{\tau=t}^{\mathcal{T}}\left(F^\top\right)^{\tau-t}Pg_\theta^{(\tau-t+1)}(d_\tau).
\]

By Definition~\ref{def:df},
  \begin{align}
        \emph{\texttt{LD}}(L_t,\psi_t^\top H\psi_t)&=\left\|\frac{\partial (\psi_t^\top H \psi_t)}{\partial \theta}-\frac{\partial (\hat\psi_t^\top H \hat\psi_t)}{\partial \theta}\right\| \notag \\
        &=\left\|2(\psi_t-\hat \psi_t)^\top H \frac{\partial  \psi_t}{\partial \theta}\right\| 
        \label{eq:LFB_0}
        \\
        &\leq 2\|H\|\left\|\psi_t-\hat \psi_t\right\|\left\|\frac{\partial  \psi_t}{\partial \theta}\right\|.\label{eq:LFD}
    \end{align}
We have used the fact that $\frac{\partial  \psi_t}{\partial \theta}=\frac{\partial  \hat\psi_t}{\partial \theta}$ to get~\eqref{eq:LFB_0}. Note that if $\mathcal{T}=T-1$, then $\psi_t=\hat \psi_t$, $\emph{\texttt{LD}}(L_t,\psi_t^\top H \psi_t)=0$. Thus, it suffices to restrict our analysis to the case when $\mathcal{T}=t+k-1$. It follows that for some $C>0$, $\rho<1$
\begin{align}
    \left\|\psi_t-\hat \psi_t\right\|&=\left\|\sum_{\tau=t+k}^{T-1}\left(F^\top\right)^{\tau-t}Pw_\tau\right\| \notag \\
        &\leq W\|P\|\sum_{\tau=t+k}^{T-1}\|F\|^{\tau-t} \label{eq:psi-hpsi_0} \\
        &\leq W\|P\| \frac{C\rho^k}{1-\rho}, \label{eq:psi-hpsi}
    \end{align}
where~\eqref{eq:psi-hpsi_0} follows from the Assumption \ref{asp:mpc} so that $\|w_t\|\leq W$ for all $t\in [T]$, and~\eqref{eq:psi-hpsi} follows from Gelfand's formula. Moreover, by Assumption \ref{asp:convex}, $\|\nabla_\theta g_{\theta_t}^{(\tau-t+1)}(d_\tau)\|\leq L$,
\begin{align}
    \left\|\frac{\partial  \psi_t}{\partial \theta}\right\|&=\left\|\sum_{\tau=t}^{t+k-1}\left(F^\top\right)^{\tau-t}P\nabla_\theta g_{\theta_t}^{(\tau-t+1)}(d_\tau)\right\| \notag\\
        &\leq L\|P\|\sum_{\tau=t}^{t+k-1}\|F\|^{\tau-t} \notag \leq L\|P\|\frac{C}{1-\rho}.\label{eq:nablapsi}
    \end{align}

Finally, combining \eqref{eq:LFD},\eqref{eq:psi-hpsi}, and~\eqref{eq:nablapsi} we have
    \begin{align}
        \emph{\texttt{LD}}(L_t,\psi_t^\top H\psi_t)\leq 2\frac{LW\|P\|^2\|H\|C^2\rho^k}{(1-\rho)^2}.
    \end{align}

\section{Proof of Theorem~\ref{thm:robustness}}
The following lemma characterizes the non-singularity of $\sum_{i = 0}^{k - 1} (F^\top)^i P$
\begin{lemma} \label{lemma:matrix_sum_non_singular}
    Under Assumption~\ref{asp:mpc}, for any finite horizon $k \geq 0$, the matrix sum $\sum_{i = 0}^{k - 1} (F^\top)^i P$ is invertible.
\end{lemma}

\begin{proof}
    Let $G_k \coloneqq \sum_{i=0}^{k - 1} (F^\top)^i$, using the algebraic identity for finite geometric series, we have that $(I - F^\top) G_k = I - (F^\top)^{k}$. Since $\rho(F) < 1$, the matrix $I - F^\top$ is invertible. Thus, we can write $G_k = (I - F^\top)^{-1} (I - (F^\top)^{k})$. The eigenvalues of $(F^\top)^{k}$ are $\lambda^{k}$ for $\lambda \in \text{spec}(F)$. Since $|\lambda| < 1$, it follows that $1 \notin \text{spec}((F^\top)^{k})$, which means $I - (F^\top)^{k}$ is invertible. As the product of two invertible matrices, $G_k$ is invertible. Finally, since $Q, R \succ 0$, we have $P$ in~\eqref{eq:dare} a positive definite matrix. Consequently, the matrix sum is invertible.
\end{proof}

Next, we start to formally prove Theorem~\ref{thm:robustness}. We immediately see that Assumption~\ref{asp:convex} holds, so both the sublinear regret bound established in Corollary~\ref{coro:final_bound} and Theorem~$2$ in~\cite{pmlr-v28-shamir13}  guarantee that the parameter sequence $\{\theta_t\}_{t \in [T]}$ converges to the minimizer of the expected loss. Let $\theta^\star = [C^\star, b^\star]^\top$ denote the limit point. Apply Assumption~\ref{asp:SGD_robustness_feasible} and denote the expected loss function as $\mathcal{L}_t(\theta) = \mathbb{E}[L_t(\theta)]$ we have:
$\nabla_{\theta} \mathcal{L}_t(\theta^\star) = 0.$
Now, we analyze the components $b$ and $C$ separately.\\
\textit{Convergence of $b$}: We calculate the gradient of $J$ with respect to $b$ and set it to zero:
\begin{equation*}
    \frac{\partial \mathcal{L}_t}{\partial b} = \mathbb{E} \left[ \frac{\partial \hat{\psi}_t(\theta)^\top H \hat{\psi}_t(\theta)}{\partial b}\right] = \mathbb{E} \left[ 2 \frac{\partial \hat{\psi}_t(\theta)}{\partial b} H \hat{\psi}_t (\theta) \right] = 0,
\end{equation*}
where
\begin{equation} \label{eq:gradient_psi_b}
     \frac{\partial \hat{\psi}_t(\theta)}{\partial b} = -\sum_{\tau = t}^{\mathcal{T}} (F^\top)^{\tau - t} P
\end{equation}
Since $H$ is positive-definite by Assumption~\ref{asp:SGD_robustness_overactuated}, and applying Lemma~\ref{lemma:matrix_sum_non_singular}, the gradient in~\eqref{eq:gradient_psi_b} is of full rank. Therefore,
\begin{equation*}
    \mathbb{E} \left[ \hat{\psi}_t (\theta)\right] = \sum_{\tau = t}^{\mathcal{T}} (F^\top)^{\tau - t} P \left( \mathbb{E}[w] - C^\star \mathbb{E}[d] - b^\star\right) = 0.
\end{equation*}
Applying Lemma~\ref{lemma:matrix_sum_non_singular} again implies that:
\begin{equation} \label{eq:optimal_b}
    b^\star = \mathbb{E}[w] - C^\star \mathbb{E}[d].
\end{equation}
\textit{Convergence of $C$}: Denote the stacked centered disturbances 
\vskip 0.2em \noindent vector $ \widetilde{\mathbf{W}}_t$ and the stacked centered embedding vector $\widetilde{\mathbf{D}}_t$ as:
\begin{equation*}
     \widetilde{\mathbf{W}}_t \coloneqq \begin{bmatrix}         \widetilde{w}_t &
        \ldots&
        \widetilde{w}_{\mathcal{T}}
     \end{bmatrix}^\top,
     \widetilde{\mathbf{D}}_t \coloneqq \begin{bmatrix}
        \widetilde{d}_t&
        \ldots&
        \widetilde{d}_{\mathcal{T}}
     \end{bmatrix}^\top,
\end{equation*}
where $\widetilde{w}_t$ and $\widetilde{d}$ are centered variables defined as:
\begin{equation} \label{eq:centered_dw}
    \widetilde{w}_t \coloneqq w_t - \mathbb{E}[w] \quad \text{and} \quad \widetilde{d}_t \coloneqq d_t - \mathbb{E}[d]
\end{equation}
Let
$
    \mathbf{M} \coloneqq \begin{bmatrix}
        P & F^\top P & \ldots & (F^\top)^k P
    \end{bmatrix}
$, function $\mathcal{L}_t$ becomes:
\begin{equation} \label{eq:expect_loss}
    \mathcal{L}_t = \mathbb{E} \left[ \left( \mathbf{M} \widetilde{\mathbf{W}}_t - \mathbf{M} \mathbf{C} \widetilde{\mathbf{D}}_t\right)^\top H \left( \mathbf{M} \widetilde{\mathbf{W}}_t - \mathbf{M} \mathbf{C} \widetilde{\mathbf{D}}_t\right) \right],
\end{equation}
where $\mathbf{C} \coloneqq I_{k \times k} \otimes C$ and $\otimes$ denotes Kronecker product. Using the linearity of expectation, \eqref{eq:expect_loss} becomes:

\begin{align*}
    & \mathbb{E} \left[  \underbracket{( \mathbf{M} \mathbf{C} \widetilde{\mathbf{D}}_t)^\top H ( \mathbf{M} \mathbf{C} \widetilde{\mathbf{D}}_t)}_{\text{Quadratic term}} \right] + \mathbb{E} \left[ \underbracket{( \mathbf{M} \widetilde{\mathbf{W}}_t  )^\top H \left( \mathbf{M} \widetilde{\mathbf{W}}_t \right )}_{\text{Constant with respect to } \mathbf{C}} \right]\\
    & -2 \mathbb{E} \left[ \underbracket{( \mathbf{M} \widetilde{\mathbf{W}}_t )^\top H ( \mathbf{M} \mathbf{C} \widetilde{\mathbf{D}}_t )}_{\text{Cross term}} \right].
\end{align*}
Since the Cross term is a scalar, the expectation of the Cross term is:
$
    \mathrm{Trace} ( \mathbf{M}^\top H \mathbf{M} \mathbf{C} \mathbb{E} [ \widetilde{\mathbf{D}}_t  \widetilde{\mathbf{W}}_t^\top]),
$
Recalling~\eqref{eq:centered_dw}, $ \mathbb{E} [ \widetilde{\mathbf{D}}_t  \widetilde{\mathbf{W}}_t^\top]$ is the cross-covariance matrix of $d$ and $w$, which is zero by assumption. Therefore, minimizing $\mathcal{L}_t$ becomes equivalent to minimizing the Quadratic term. Similarly, the expectation of the quadratic term can be written as:
$\mathrm{Trace} ( ( \mathbf{M} \mathbf{C} )^\top H ( \mathbf{M} \mathbf{C}) \mathbb{E} [\widetilde{\mathbf{D}}_t \widetilde{\mathbf{D}}^\top_t ]),$
where $\mathbb{E} [\widetilde{\mathbf{D}}_t \widetilde{\mathbf{D}}^\top_t ]$ is the variance of $\widetilde{\mathbf{D}}_t$ since the expectation of $\widetilde{\mathbf{D}}_t$ is zero by~\eqref{eq:centered_dw}. Therefore, due to the positive-definiteness of $H$, minimizing the Quadratic term is equivalent to setting $\mathbf{M} \mathbf{C}$ to zero, whose necessary condition is:
$
   \sum_{\tau = t}^{\mathcal{T}} (F^\top)^{\tau - t} P C = 0,
$
which directly leads to $C = 0$ by using Lemma~\ref{lemma:matrix_sum_non_singular}. Substituting $C = 0$ into~\eqref{eq:optimal_b} yields $b = \mathbb{E}[w]$.

\end{document}